\documentstyle[graphicx,multicol,prl,aps,epsf]{revtex}
\begin{document}
\title{Tunneling through an Anderson impurity between superconductors}
\author{Yshai Avishai 
\cite{email}
$^1$, Anatoly Golub$^1$ and Andrei D. Zaikin$^{2,3}$}
\address{$^1$ Department of Physics, Ben-Gurion University of the Negev,
Beer-Sheva, Israel\\
$^2$ Forschungszentrum Karlsruhe, Institut f\"ur Nanotechnologie,
76021 Karlsruhe, Germany\\
$^3$ I.E.Tamm Department of Theoretical Physics, P.N.Lebedev
Physics Institute, 117924 Moscow, Russia}

\date{\today}
\maketitle
 
\begin{abstract} 
We consider an Anderson impurity ($A$)
weakly connected to a superconducting electrode ($S$) on one side 
and a superconducting or a normal metal electrode ($N$) on the other side. 
A general path integral formalism is developed and the response of 
$SAN$ and $SAS$ junctions 
to a constant voltage bias $V$ is elucidated, 
using a combination of the Keldysh 
technique (to handle non-equilibrium effects) and 
a dynamical mean field approximation 
(to handle repulsive Hubbard interactions). 
An interesting physics is exposed at sub-gap voltages 
($eV < \Delta$ for $SAN$ and $eV < 2 \Delta$ for $SAS$).
For an $SAN$ junction, Andreev reflection is strongly
affected by Coulomb interaction. 
For superconductors 
with $p$-wave symmetry the junction conductance exhibits a
remarkable peak at $eV < \Delta$, while for
superconductors with $s$-wave symmetric
pair potential the peak is shifted 
towards the gap edge $eV=\Delta$ and 
strongly suppressed
if the Hubbard repulsive 
interaction increases.
Electron transport in $SAS$ junctions is determined by an
interplay between multiple Andreev reflection (MAR) and Coulomb effects.
For $s$-wave superconductors the usual peaks in the conductance that 
originate from MAR  are shifted by interaction 
to larger values of $V$. They are also 
suppressed as the Hubbard interaction 
strength grows.  For $p$-wave superconductors the sub-gap
current is much larger and the $I-V$ characteristics reveal 
a new feature, namely, a peak in the current resulting from
a mid-gap bound state in the junction.
\end{abstract}
\begin{multicols}{2}

\noindent
\narrowtext
\section 
 {Introduction}

The dynamical behavior of Josephson junction strongly 
depends, among other factors, 
on its transparency. If the insulating
barrier is not too high then the concept 
of nonlinear tunneling becomes
relevant. In this case
the characteristic dynamical conductance $dI/dV$
at applied voltages $V$ less than the
superconducting gap $\Delta$ shows a sub-gap
structure. An explanation of this behavior 
was given some time
ago \cite{tin,arnold}, based on the mechanisms of multiple
Andreev reflections (MAR). Recently, the sub-gap current was
calculated for the case of electron
tunneling through a junction with resonant impurity
\cite{gol}. Rapid progress in the technology of 
superconducting junctions makes it possible to fabricate junctions 
composed of quantum dots weakly coupled to superconducting or 
normal electrodes. The basic physics of 
such a device can be elucidated once it is modeled as an 
Anderson impurity center.
In this case 
the Coulomb interaction is expected to strongly affect the
 tunneling current in general and the sub-gap current in particular.
Since the sub-gap current is originated from multiple Andreev 
reflections,
 its physics has a close similarity to that of
the Josephson current. In this context, it is established \cite{glaz} 
that the tunneling through a quantum dot is suppressed if the
effective Kondo temperature 
$T_{K}$ =$\sqrt{U\Gamma}\exp{[-\pi|\epsilon_{0}|/2\Gamma]}$ 
is small as compared to the superconducting gap $\Delta$. Hereafter, 
$U$ is the Hubbard repulsion strength, 
$\epsilon_{0}$ is the orbital energy of the dot electron and $\Gamma$ 
is the width of this energy state. Strong interaction-induced suppression 
of the current through superconducting quantum dots was also
observed experimentally \cite{RBT}.

Quite recently detailed measurements of the $I-V$ curves in
atomic-size metallic contacts were performed \cite{scheer}.
An explanation of the observed $I-V$
curves were given \cite{cue} in terms of 
the atomic valence orbitals which represent different conducting channels. The
Coulomb interaction was considered there
to be screened as in bulk metals. However, for quantum dots and
break junctions the screening is virtually 
ineffective and an unscreened Hubbard-type repulsive interaction 
emerges. In this case the Kondo temperature $T_{K}$ becomes a relevant parameter,
separating levels with $T_{K}> \Delta$ (which are responsible for high, nearly
resonant conductance) from levels with $T_{K}< \Delta$, 
in which the conductance is strongly influenced by interaction.
 
One of the main goals of the present paper is to develop a detailed
theoretical analysis of an interplay between the phenomena of 
multiple Andreev reflections and Coulomb interaction in superconducting
quantum dots. Even though both MAR and Coulomb effects have
been intensively studied in the literature over past decades, an 
interplay between them -- to the best of our knowledge -- was not
elucidated until now. In this paper we will demonstrate that these two phenomena,
being combined in superconducting quantum dots, lead to novel
physical effects and -- depending on parameters -- may dramatically
influence the sub-gap conductance pattern of the system. In short,
Coulomb suppression of MAR turns out to be much more pronounced than, say, that
of single electron tunneling. This is because during MAR cycles
at relatively low voltages the charge is transferred by large quanta, much
larger than the electron charge $e$.

Another important issue in the study of $SAS$ and $SAN$
junctions is the parity of the order parameter 
of the superconducting electrodes. For example,  
the order parameter of the recently discovered \cite{maeno} superconducting material
 $Sr_{2}RuO_{4}$ 
is believed to have a $p$-wave symmetry \cite{sig}. If a superconductor of this type
is properly oriented with respect to the tunneling
direction  the principal contribution to the Josephson current
comes from a bound state \cite{pwave,Hu} formed at the 
contact point. This bound state arises since
the pair potential has an opposite sign for injected and 
reflected quasiparticles and 
is expected  to play an important role  
in the formation of sub-gap currents.

In the present work we expose the physics of $SAS$ 
and $SAN$ junctions subject to a finite potential bias. 
In particular, we calculate the tunneling
current and the dynamical conductance for junctions 
consisting of $s$- and $p$-wave superconductors. The main 
steps required for treating the pertinent 
many-body problem can be summarized as follows: 1) Taking 
the Fermi energy of the unbiased lead as an energy reference, 
the site energy $\epsilon_{0}$ of 
the Anderson impurity is chosen such that $\epsilon_{0} <0$ 
while $U+\epsilon_{0} > 0$. These inequalities assure that assuming the quantum dot 
to be at most singly occupied should be an excellent approximation.
2) To handle the strong
interaction appearing in the Hubbard term, a mean field
approximation \cite{Arovas,AG} is adopted.  
3) The formalism should take into account 
the nonequilibrium nature of the physical system.
For this purpose, the standard approach 
is to start from the expression for the kernel of the
evolution operator or the generating functional, 
which is the analog of the partition function
in the equilibrium case, evaluated, however, on a Keldysh contour \cite{Keldysh}
(see review article in Ref. \onlinecite{schon}). At the end of this procedure  
one is able to calculate the $SAN$ Andreev conductance analytically, 
and to get expressions for the non-linear response of $SAS$ junctions 
which are amenable for numerical evaluation.
 
The technical procedure by which we manage 
to advance the calculations is detailed below
in section 2, where we derive an effective action for $SAS$ and $SAN$
junctions. In section 3 we discuss the dynamical mean field approximation
adopted in the present work in order to treat interaction effects. 
Concrete results pertaining to sub-gap current in $SAS$ junctions 
and differential conductance in $SAN$ junctions are presented 
and discussed in section 4. The paper is then concluded 
and summarized in section 5. Some technical details of the calculation 
are given in Appendix. 
\section{General Analysis}

\subsection{The Model}

Consider a system consisting of two superconducting
wide strips on the left ($x<0, -\infty < y < \infty$) 
and on the right ($x>0, -\infty < y < \infty$) weakly connected 
by a quantum dot through which an electron 
tunneling takes place. This system can be described by the Hamiltonian  
\begin{eqnarray}& &
\bbox{H}=\bbox{H}_{L}+\bbox{H}_{R}+ \bbox{H}_{\rm dot}+ 
\bbox{H}_{\rm t}.
\label{tot}
\end{eqnarray}
The Hamiltonians of the left and right superconducting 
electrodes have the standard BCS form
\begin{eqnarray} & &
\bbox{H}_j=\int dr [\Psi^{\dagger}_{j 
\sigma}(\bbox{r})\xi({\bf \nabla})\Psi_{j \sigma}(\bbox{r}) \nonumber \\ 
&& -\lambda \Psi^{\dagger}_{j \uparrow}
(\bbox{r})\Psi^{\dagger}_{j \downarrow}(\bbox{r})
	\Psi_{j \downarrow}(\bbox{r})\Psi_{j \uparrow}(\bbox{r})].
\label{BCS}
\end{eqnarray}
Here $\Psi^{\dagger}_{j \sigma}$ ($\Psi_{j \sigma}$) are the electron creation
(annihilation) operators, $\lambda$ is the BCS coupling constant,
$\xi ({\bf \nabla})=-{\bf \nabla}^2/2m-\mu$ and $j=L,R$. Here and below we 
set the Planck's constant $\hbar =1$. Whenever appropriate,
the spin, space and time dependence of all the field operators 
will not be explicitly displayed.

The quantum dot is treated as an 
Anderson impurity center located at $x=y=0$. It is described by 
the Hamiltonian
\begin{eqnarray} & &
\bbox{H}_{\rm dot}= \epsilon_0\sum_{\sigma }C^{\dagger}_{\sigma }C_{\sigma }+
UC^{\dagger}_{\uparrow }C_{\uparrow} C^{\dagger}_{\downarrow}C_{\downarrow}, 
\label{dot}
\end{eqnarray}
where $C^{\dagger}_{\sigma }$ 
and $C_{\sigma }$ are the electron operators in the dot.
The impurity site energy $\epsilon_0$ (counted from the Fermi energy $\mu$) 
is assumed to be far below the Fermi level $\epsilon_0<0$. The presence of a
strong Coulomb repulsion $U> -\epsilon_0$ 
between electrons in the same orbital
guarantees that the dot is at most singly occupied. 

Electron tunneling through the dot is accounted for by means of the term,
\begin{eqnarray} & &
\bbox{H}_{\rm t} = \sum_{j=L,R}{\cal T}_j\sum_{\sigma}\Psi^{\dagger}_{j \sigma}(0)C_{\sigma} + {\rm h.c.},
\label{int} \end{eqnarray} 
where ${\cal T}_{L(R)}$ are the effective transfer amplitudes between the left
(right) electrode and the dot.

In what follows we will always assume that, if a bias voltage 
$V$ is applied to the system from, say, right to left, 
the entire voltage drop occurs across the
dot. Hence, the quasiparticle distribution functions in
the leads are the Fermi ones, with the chemical potentials of the 
electrodes shifted with respect to each other by $eV$. 

\subsection{Evolution Operator}

Complete information about the quantum dynamics of the system is
contained within the evolution operator defined 
on the Keldysh contour \cite{Keldysh} $K$ (which consists of 
forward and backward oriented time branches). The
kernel $J$ of this evolution operator can be expressed in terms of 
a path integral, 
\begin{eqnarray} & &
J=\int {\cal D} \bar{\Psi}{\cal D}\Psi {\cal D}\bar{C}{\cal D}C\exp(iS),
\label{pathint}
\end{eqnarray}
over the fermion fields corresponding to the operators 
$\Psi^{\dagger}$, $\Psi$, $C^{\dagger}$ and $C$
(here the field $\bar{\Psi}$ corresponds to $(\Psi_{L \uparrow}^{\dagger},
\Psi_{L \downarrow}^{\dagger},\Psi_{R \uparrow}^{\dagger},
\Psi_{R \downarrow}^{\dagger})$ and similarly for other fields), 
$S=\int_{K}Ldt$ is the action and $L$ is the 
Lagrangian pertaining to the Hamiltonian (\ref{tot}). The external 
fields (e.g. electromagnetic fields) 
can be treated as the source terms for the action, though, the
fluctuating parts of these fields should be integrated as well.

Usually it is convenient to perform an operator rotation 
$C \to c$ and $\Psi \to \psi$ in Keldysh space: 
\begin{eqnarray} & &
\bar{c}=\bar{C}\sigma_z\hat Q^{-1}, \;\;\; c=\hat QC;\;\;\;
\bar{\psi}=\bar{\Psi}\sigma_z\hat Q^{-1}, \;\;\; \psi=\hat Q\Psi .
\label{transform}
\end{eqnarray}  
Here $\sigma_z$ is one of the Pauli matrices $\sigma_x$, $\sigma_y$, $\sigma_z$ 
operating in the Keldysh space and 
\begin{eqnarray}
\hat Q&=\frac{1}{\sqrt{2}}&\left ( \begin{array}{cc}
1&-1\\
1&1
\end{array}\right) 
\label{Qmatrix}
\end{eqnarray}  
is the Keldysh matrix. The Grassman 
variables $\bar{c}$, $c$, $\bar{\psi}$,
$\psi$ are now defined solely on the forward time branch. 

The transformation of the Green functions follows directly from (\ref{transform}).
One starts from the $2\times2$ matrix $\hat{\tilde{G}}$ of the Green functions
defined in terms of the initial electron operators. The elements of the 
matrix $\hat{\tilde{G}}$ are the Green functions $\hat{\tilde{G}}_{ij}$ 
with $i,j=+,-$ according 
to whether the time belongs to the upper 
or the lower branch of the Keldysh contour K. Of these four 
Green functions only three are independent. Under the operator rotation 
(\ref{transform}) the Green-Keldysh matrix $\hat{\tilde{G}}$ is
transformed as $\hat{G}=Q^{-1} \hat{\tilde{G}}Q$, where 
\begin{eqnarray}
\hat{G}=\left ( \begin{array}{cc}
\hat{G}^{R}&\hat{G}^{K}\\
\hat{0}&\hat{G}^{A}
\end{array}\right)
\label{Eq_GKeld}
\end{eqnarray}
and
\begin{eqnarray}
	&& \hat{G}^{R}=-i\theta(t-t')\left<\psi(r,t)\psi^{\dagger}(r',t')+
	\psi^{\dagger}(r',t')\psi(r,t) \right>, \nonumber \\
	&&G^{A}=i\theta(t'-t)\left<\psi(r,t)\psi^{\dagger}(r',t')+
	\psi^{\dagger}(r',t')\psi(r,t) \right>, \nonumber \\ 
	&&G^{K}=-i\left<\psi(r,t)\psi^{\dagger}(r',t')+
	\psi^{\dagger}(r',t')\psi(r,t) \right>, 
	\label{Eq_GK}
\end{eqnarray}
are respectively retarded, advanced and Keldysh  Green functions.
Each of these matrices is in turn $2\times2$ matrix in the Nambu space. 

The path integral (\ref{pathint}) is now expressed in terms of the new Grassman 
variables 
\begin{eqnarray} & &
J=\int {\cal D} \bar{\psi}{\cal D}\psi {\cal D}\bar{c}{\cal D}c
\exp(iS_{\rm dot}+iS_0[\bar{\psi},\psi]),
\label{pathint2}
\end{eqnarray}
where
\begin{eqnarray} & &
S_{\rm dot}=\int dt\left[\bar{c}\left(i\frac{\partial}{\partial t}-\tilde{\epsilon}
\tau_{z}\right)c+\frac{U}{2}(\bar{c}c)^2\right], 
\label{sdot}
\end{eqnarray}
\begin{eqnarray} & &
S_0=\int dt \sum_{j=L,R}\bigg[\int_jdr\bar{\psi}_{j}(r,t)
\hat G_{j}^{-1}\psi_{j}(r,t) \nonumber \\
&& +({\cal T}\bar{\psi}_{j}(0,t)\tau_{z}c(t)+{\rm c.c.})\bigg].
\label{SS}
\end{eqnarray}
Here we defined $\tilde{\epsilon}=\epsilon_{0}+U/2$. 
In order to obtain the expression for the operator $\hat G_{j}^{-1}$ we
employ the standard Hubbard-Stratonovich transformation of the $\Psi^4$-term
in (\ref{BCS}) and introduce additional path integrals over the complex scalar 
order parameter field $\Delta (r,t)$ defined on the Keldysh contour, see 
e.g. Ref. \onlinecite{schon}. Since here we are not interested in the fluctuation 
effects for the order parameter field, we will evaluate the
path integral over $\Delta$ by means of the saddle point approximation,
which amounts to setting $\Delta (r,t)$ equal to the equilibrium
superconducting order parameter values $\Delta_{L,R}$ of the left and the right
electrodes. If needed, fluctuations
of the order parameter field (both the amplitude and the phase) can easily be 
included into our consideration along the same lines as it was done in Ref. 
\onlinecite{schon}. Disregarding such fluctuations here, we find   
\begin{eqnarray} & & 
\hat G_{L,R}^{-1}(\xi )=i\frac{\partial}{\partial t}-
\tau_{z}\xi({\bf \nabla})+\tau_+\Delta_{L,R}+\tau_-\Delta^*_{L,R},
\label{G-1}
\end{eqnarray}
where we define $\tau_{\pm}=(\tau_x\pm i\tau_y)/2$. Here and below $\tau_{x},
\tau_{y},\tau_{z}$ is the set of Pauli matrices operating in the Nambu space
(for the sake of clarity we chose a different notation from that used for
Pauli matrices operating in the Keldysh space).

\subsection{Effective Action}

Let us now proceed with the derivation of the effective action for our model.
We first notice that the $\psi$-fields dependent part $S_0$ of the total action
is quadratic in these fields. Hence, the integrals over $\bar \psi$ and $\psi$
in (\ref{pathint2}) can be evaluated exactly, resulting in an action 
$S_{\rm env}(\bar{c},c)$ defined as,
\begin{eqnarray} & & 
\exp (iS_{\rm env}[\bar{c},c])=
\int {\cal D} \bar{\psi}{\cal D}\psi 
\exp(iS_0[\bar{\psi},\psi]).
\label{pathint4}
\end{eqnarray}
Its physical content can be understood as follows;
One can say that electrons in two superconducting bulks serve as an
effective environment for the quantum dot. 
Integrating out these electron variables in the spirit of the 
Feynman-Vernon influence functional approach \cite{FV} one arrives
at the ``environment'' contribution to the action $S_{\rm env}$
expressed only in terms of the Anderson impurity variables $\bar c$ and $c$.

Due to the fact that coupling to the leads is concentrated at one 
point $(x,y)=(0,0)$ we can integrate out the fields inside the 
superconductors (hereafter referred  as bulk fields) 
and obtain an effective action in terms of 
fermion operators with arguments solely on the surface.
In order to achieve this central goal let us first 
note that translation invariance along $y$ 
permits the Fourier-transform in eq. (\ref{SS}) in this direction. 
The problem then reduces to a one dimensional one 
with fermion fields $\psi_{k}(x)$ where $k$ is the momentum 
along $y$. Gaussian integration over the bulk fields 
can be done with the help of the saddle point method. 

Let us consider, say, the left superconductor and omit
the subscript $j=L$ for the moment. The 
pertinent equation for the optimal field reads,
\begin{eqnarray} & &
\hat G^{-1}(\xi_x) \tilde \psi_{k}(x)=0,
\label{Eq_SP}
\end{eqnarray} 
where $\xi_x= -(1/2m)(\partial^2/\partial x^2)-\mu_k$ and $\mu_k=\mu -k^2/2m$. 

Let us decompose
$\tilde \psi_{k}(x)=\psi^{b}_{k}(x)+\psi_{k}(0)$ in such a way that 
on the surface one has $\psi^{b}_{k}(0)=0$.
The bulk field  $\psi^{b}_{k}(x)$ satisfies the
inhomogeneous equation:
\begin{eqnarray} & &
\hat G^{-1} (\xi_x) \psi^{b}_{k}(x)=-\mu_{k} \tau_{z}\psi_{k}(0). %
\label{Eq_In}
\end{eqnarray}
In the right hand side of this equation
we employed the standard quasiclassical (Andreev) approximation
which makes use of the fact that the superconducting gap as well
as other typical energies of the problem are all much smaller than the 
Fermi energy.
 
In order to solve eq. (\ref{Eq_In}) we find the Green function 
$\hat G_{k}(x,t;x',t')$ (which satisfies the same equation albeit with a 
$\delta$ -function on the right hand side) and require 
$\hat G$ to vanish at $x=0$. The solution of eq. (\ref{Eq_In})
is then exploited to express
$\psi^{b}_{k}(x)$ in terms of the surface fields $\psi_{k}(0)$.
Combining the result with eq. (\ref{SS}) 
we arrive at the intermediate effective action $\tilde S$ 
for a superconducting electrode which depends on the $\psi$-fields at the surface,
\begin {eqnarray} & &
\tilde S=i\int dt \int dt'\sum_{k}\frac {v_{x}}{2}\bar{\psi}_{k}(0,t)
 \tau_{z}\hat g(t,t')\psi_{k}(0,t'),
\label{SL}
\end{eqnarray}
where $v_x=\sqrt{2 \mu_{k}/m}$ is the quasiparticle velocity in 
the $x$-direction.
For a uniform superconducting half-space (here the left one), 
the Green-Keldysh matrix 
\begin{eqnarray} & &
\hat g(t,t')\tau_z= -i \frac{v_{x}}{2}
\frac{\partial}{\partial x} \int_{L} dx'\hat G_{k}(x,t;x',t')|_{x=0}
\label{Eq_Seff}
\end{eqnarray}
(which has the structure (\ref{Eq_GKeld})) is expressed in terms of
the Eilenberger functions \cite{Eil} as follows:
\begin{eqnarray} & &
\hat g(t,t')=e^{\frac{i\varphi(t)\tau_z}{2}}\int \hat g(\epsilon
)e^{-i\epsilon (t-t')}
\frac{d\epsilon}{2\pi}e^{-\frac{i\varphi(t')\tau_z}{2}},
\label{gL1}
\end{eqnarray} 
where 
\begin{eqnarray} & &
\hat g^{R/A}(\epsilon )= \frac{(\epsilon \pm i0)\tau_z +i|\Delta|\tau_y}
{\sqrt{(\epsilon \pm i0)^2-|\Delta|^2}},
\label{gL2}
\end{eqnarray}
\begin{eqnarray} & &
\hat g^K(\epsilon )=(\hat g^R(\epsilon )-\hat g^A(\epsilon ))
\tanh (\epsilon /2T).
\label{gL3}
\end{eqnarray} 
Here $\varphi (t)=\varphi_0 
+2e\int^tV(t_1)dt_1$ is the time-dependent phase of
the superconducting order parameter and $V(t)$ is the electric potential of
the superconducting electrode.

An identical procedure applies for the right electrode.
Each superconductor is thus described by a zero-dimensional
action, respectively $\tilde S_{L}$ and $\tilde S_{R}$, coupled by an 
on-site hopping term with the Anderson impurity.  
It is now possible to integrate out these surface fields. The integral
\begin{eqnarray} & & 
\tilde J=\int {\cal D} \bar{\psi} (0){\cal D} \psi (0) 
\exp (i\tilde S_{L,R} \nonumber \\
&& +i\int dt({\cal T}_{L,R}\bar{\psi_k}(0)\tau_zc +{\rm c.c.}))
\label{pathint3}
\end{eqnarray}
can easily be evaluated, so that the contribution of the
superconductors to the total effective action of our model is manifested 
in $S_{\rm env }$ defined as 
\begin{eqnarray} & &
S_{\rm env}=2i\int dt \int
dt'\sum_k\bar{c}(t)\tau_z[\frac{{\cal T}_{L}^2}{v_x}
\hat g_L(t,t') \nonumber \\
&& +\frac{{\cal T}_{R}^2}{v_x}\hat g_R(t,t')]c(t').
\label{Eq_Sb0}
\end{eqnarray}
Note that in deriving (\ref{Eq_Sb0}) we made use of the normalization condition
\cite{Eil} $\hat g_{L,R}^2=1$. 

Eq. (\ref{Eq_Sb0}) is valid for an arbitrary pairing symmetry. In the
case of unconventional superconductors the Green functions $\hat g_{L,R}$
depend explicitly on the direction of the Fermi velocity. For uniform
$s$-wave superconductors such dependence is absent and eq. (\ref{Eq_Sb0}) 
can be simplified further. Defining the tunneling rates between the left (right) superconductor and the dot as
\begin{eqnarray} & & 
\Gamma_{L(R)} = 4\sum_k \frac{{\cal T}_{L(R)}^2}{v_x},
\label{rates}
\end{eqnarray}
we obtain
\begin{eqnarray} & &
S_{\rm env}=\frac{i}{2}\int dt \int
dt'\bar{c}(t)\tau_z[\Gamma_L\hat g_L(t,t')+\Gamma_R\hat g_R(t,t')]c(t').
\label{Eq_Sb}
\end{eqnarray}
The definition (\ref{rates}) requires a comment. Here we consider the situation
with one conducting channel in the dot. In this case the transfer
amplitudes ${\cal T}_{L,R}$ should effectively differ from zero only for 
$|v_x| \approx v_F$. One can easily generalize the 
action (\ref{Eq_Sb}) to the situation with several or even many conducting
channels. In this case the summation over momentum (essentially equivalent 
to the summation over conducting modes) should be done in (\ref{Eq_Sb})
and some other dependence of ${\cal T}^2_{L,R}$ on $v_x$ should apply.
For instance, for tunnel junctions in the many channel limit
one can demonstrate \cite{zaikin} that ${\cal T}^2_{L,R} \propto v_x^3$.
It is also quite clear that the transfer amplitudes ${\cal T}_{L,R}$
{\it cannot} be considered as constants independent of the Fermi velocity
direction, as it is sometimes assumed in the literature. In that case 
the sum (\ref{rates}) would simply diverge at small $v_x$ in a clear 
contradiction with the fact that quasiparticles with $v_x \to 0$
should not contribute to the current at all. This ``paradox'' is resolved
in a trivial way: the amplitudes ${\cal T}_{L,R}$ do depend on $v_x$ 
and, moreover, they should vanish at $v_x \to 0$. For
further discussion of this point we refer the reader to Ref. \onlinecite{zaikin}.

Combining eqs. (\ref{pathint2}) and (\ref{pathint4}) we arrive at the expression 
for the kernel of the evolution operator $J$ solely in terms of the 
fields $\bar{c}$ and $c$: 
\begin{eqnarray} & &
J=\int {\cal D} \bar{c}{\cal D}c \exp(iS_{\rm eff}),\;\;\;\;\;
S_{\rm eff}[\bar{c},c]=S_{\rm dot}+S_{\rm env}.
\label{Jeff}
\end{eqnarray}
Here $S_{\rm eff}[\bar{c},c]$ (defined by eqs. (\ref{sdot}) and (\ref{Eq_Sb})) 
represents the effective action for a quantum dot between two superconductors. 

\subsection{Transport current}

In order to complete our general analysis let us express the current through
the dot in terms of the correlation function for the variables
$\bar{c}$ and $c$. This goal can be achieved by various means. For instance,
one can treat the superconducting phase difference across the dot as a 
source field in
the effective action and obtain the expression for the current just by
varying the corresponding generating functional with respect to this phase
difference. Another possible procedure is to directly employ the 
general expression for the current in
terms of the Green-Keldysh functions of one (e.g. the left) superconductor, 
with arguments at the impurity site:
\begin{eqnarray} & &
I=\frac{e}{4m}\int dy(\partial_x
-\partial_{x'}){\rm Tr}[\hat G^K(xy,x'y';t)]_{x=x'}, 
\label{Eq_J}
\end{eqnarray}
where the trace is taken in the Nambu space.

As before, it is convenient to split the bulk and the surface variables.
After a simple algebra we transform eq. (\ref{Eq_J}) into the following result:
\begin{eqnarray} & &
I=-i\frac{e}{4}\sum_{k}v_{x}{\rm Tr}[\hat g_L \hat G_{\psi}-
{\rm h.c.}]|_K ,
\label{Eq_JG}
\end{eqnarray}
where $G_{\psi}=-i<\psi_{k}(0)\bar{\psi}_{k}(0)>$ is 
the Green-Keldysh function for the surface $\psi$-fields.
Here and below the  integration over the internal time variables in the product
of matrices is implied and $(...)|_{K}$ means the Keldysh component of this
product. 

Finally, let us express the function $\hat G_{\psi}$
in terms of the correlator for the fields $\bar{c}$ and $c$. 
Let us consider the generating functional for the surface fields
\begin{eqnarray} & &
Z[\bar \eta , \eta ]= \tilde J [{\cal T}_L\bar\tau_z c +\bar \eta,
{\cal T}_L\tau_z c + \eta],
\label{genfl}
\end{eqnarray}
where the path integral $\tilde J$ is defined in (\ref{pathint3}). The 
functional derivative of (\ref{genfl}) with respect to the $\eta$-fields 
just yields the function $\hat G_{\psi}$:
\begin{eqnarray} & &
\hat G_{\psi}= i\frac{\delta^2Z}{\delta \bar \eta \delta \eta}
|_{\bar \eta =\eta =0}.
\label{defin}
\end{eqnarray}
Evaluating the path integral (\ref{genfl}) and making use of (\ref{defin})
we arrive at the following identity
\begin{eqnarray} & &
i\hat G_{\psi}=-\frac{2}{v_{x}}\hat g_L\tau_z+
\frac{4{\cal T}_L^2}{v_{x}^{2}}\langle c\bar{c} \rangle .
\label{identity}
\end{eqnarray}

Combining (\ref{Eq_JG}) and (\ref{identity}) with the 
condition $\hat g_L^2=1$ we observe that the
contribution of the first term in the right-hand side of (\ref{identity})
to the current vanishes identically, and only the second term 
$\propto \langle \bar cc \rangle$ turns out to be important. Making use
of the definition (\ref{rates}) and symmetrizing the final result with respect to $R$ and $L$ we arrive at the following expression for the current
\begin{eqnarray} & &
I=\frac{e}{8}{\rm Tr}
[(\Gamma_L\hat g_L-\Gamma_R\hat g_R)\langle \bar{c}c\rangle |_K + {\rm h.c.}].
\label{tok}
\end{eqnarray}
This expression completes our derivation. We have demonstrated that 
in order to calculate
the current through an interacting quantum dot between two superconducting
electrodes it is sufficient to evaluate the correlator 
$\langle \bar{c}c\rangle$ in the model defined by the effective
action $S_{\rm eff}=S_{\rm dot}+S_{\rm env}$ (\ref{sdot}), (\ref{Eq_Sb}). 
Our approach enables one to investigate both equilibrium and 
nonequilibrium electron transport in superconducting quantum dots. In the 
noninteracting limit $U \to 0$ the problem reduces to a Gaussian one.
In this case it can easily be solved and, as we will demonstrate below,
the well known results describing normal and superconducting contacts 
without interaction can be re derived in a straightforward manner. In
the interacting case $U \neq 0$ the solution of the problem involve
approximations. One of them, the dynamical mean field approximation,
is described in the next section.

\section{Mean Field Approximation}

In order to proceed further let us 
decouple the interacting term in (\ref{sdot}) by means
of a Hubbard-Stratonovich transformation \cite{Arovas,AG}
introducing additional scalar fields $\gamma_{\pm}$. The kernel $J$
now reads,
\begin{eqnarray} & &
J= \nonumber \\
&& \int {\cal D} \bar{c} {\cal D} c 
{\cal D} \gamma_{+} {\cal D}\gamma_{-} 
\exp \left[iS[\gamma ]+i\int dt\bar{c}\left(i\frac{\partial}{\partial t}-\tilde{\epsilon}
\tau_{z}\right)c\right],
\label{HS}
\end{eqnarray}
\begin{eqnarray} & &
S[\gamma ]=\int dt\left(\bar{c}\gamma_{+}\sigma_{x}c+\bar{c}\gamma_{-}c 
-\frac{2}{U}\gamma_{+}\gamma_{-}\right).
\label{Seff1}
\end{eqnarray}
These equations are still exact. Now let us assume that the effective Kondo 
temperature 
$T_{K}$ =$\sqrt{U\Gamma}\exp{[-\pi|\epsilon_{0}|/2\Gamma]}$
is smaller than the superconducting gap $\Delta$. In this case 
interactions can be accounted for within
the mean field (MF) approximation. The fields 
$\gamma_{\pm}$ in (\ref{Seff1}) can be determined 
from the saddle point conditions 
\begin{eqnarray} & &
\delta J/\delta \gamma_{\pm}=0.
\label{dJ}
\end{eqnarray}
In general these two equations contain an explicit dependence on the time
variable. Let us
average these equations over time 
and consider $\gamma_{\pm}$ as time independent parameters. This approximation
is equivalent to retaining only the first moment of $\gamma_{\pm}$.
The self-consistency equations (\ref{dJ}) now read
\begin{eqnarray}
\gamma_{+}=\frac{U}{2}\int dt<\bar{c}c>,\\ 
\gamma_{-}=\frac{U}{2}\int dt<\bar{c}\sigma_{x}c>.
\label{Eq_gamma}
 \end{eqnarray}
As it turns out from our numerical analysis (to be described below), 
the parameter $\gamma_{+}$
has a negligible effect on the sub-gap current. It just slightly
renormalizes the coupling constants ${\cal T}_{L,R}$ of our model.
On the other hand, the second parameter,
$\gamma_{-}$, has a large impact on the 
$I-V$ characteristics. Therefore in what
follows we will set $\gamma_{+}=0$ and take into account only the 
second self-consistency equation (\ref{Eq_gamma}) for $\gamma_{-}$. 
Under this approximation the effective action of our model acquires
the following form
\begin{eqnarray} &&
S_{\rm eff} [\gamma ] = \int \frac{d\epsilon}{2\pi} \int
d\epsilon'\bar{c} \hat M(\epsilon,\epsilon')c, \nonumber \\
&& \hat M(\epsilon,\epsilon') = \delta
(\epsilon-\epsilon')[\epsilon +\gamma_{-}-\tau_z\tilde{\epsilon}
+i\tau_z(\Gamma_R/2) \hat g_{R}(\epsilon)] \nonumber \\
&& +i\tau_z(\Gamma_L/2)\hat g_{L}(\epsilon,\epsilon'). 
\label{Seff2}
\end{eqnarray}
Here and below we deliberately choose the electrostatic potential of the right
electrode to be equal to zero. In this case the Keldysh matrix $\hat g_R$ is
diagonal in the energy space. 
Performing the functional integration over Grassman variables $\bar c$ and $c$
we can cast the self-consistency equation
(\ref{Eq_gamma}) for $\gamma_{-}$ in terms of the matrix     
\begin{eqnarray}
\hat M^{-1}=\left ( \begin{array}{cc}
(\hat M^R)^{-1}&-(\hat M^R)^{-1}\hat M^K(\hat M^A)^{-1}\\
\hat 0 & (\hat M^A)^{-1} 
\end{array}\right), 
\label{Eq_M}
\end{eqnarray}  
where $\hat M^{R}$, $\hat M^{A}$  $\hat M^{K}$ are three independent
elements of the Keldysh matrix $\hat M$ (\ref{Seff2}). Recall that each of 
these elements 
is a $2\times2$ matrix in the Nambu space and an infinite matrix in the energy
space. Eq. (\ref{Eq_gamma}) for $ \gamma_{-}$ can now be rewritten as 
\begin{eqnarray} & &
\gamma_{-}=i\frac{U}{2}{\rm Tr}(\hat M^R)^{-1}\hat M^K(\hat M^A)^{-1},
\label{samosogl}
\end{eqnarray} 
with the trace being taken both in the energy and spin spaces.

 Finally, 
employing the MF approximation 
for the Hubbard interaction 
as was implied in the calculation of
$\gamma_{-}$, we get the current 
as a difference of symmetric forms, 
\begin{eqnarray}
I=\frac{e\Gamma_L\Gamma_R}{8} {\rm Tr}
[(\hat N_{L}
\hat g_{R}^{R}€-
(L\leftrightarrow R))+{\rm h.c.}],\label{Eq_Jfinal} \\
\hat N_{L,R}=(\hat M^{R})^{-1}\tau_z\hat g^{K}_{L,R}(\hat M^{A})^{-1}.\nonumber
\end{eqnarray}

Consider now the case of a constant in time voltage bias $V$ and recall that 
the entire voltage drop occurs across the
quantum dot. Setting the phase of the right electrode equal to
zero, for the phase of the left superconductor we obtain 
$ \varphi(t)=2eVt +\varphi_{0}$. Let us express $\hat g_L$ in
terms of the matrix elements in the energy space \cite{arnold}
\begin {eqnarray} &&
(\epsilon |\hat g_{L}|\epsilon')=\sum_{s=0,\pm 1} 
\delta(\epsilon -\epsilon'+2seV)\hat g_{L}(\epsilon ,\epsilon +2seV),
\label{Eq_GE} \nonumber \\
&& \hat g_{L}(\epsilon ,\epsilon +2seV)=(\hat g^{11}_{L}(E-eV)P_{+}+
g^{22}_{L}(\epsilon +eV)P_-)\delta_{0,s} \nonumber \\
&& +e^{i\varphi_{0}}g^{12}_{L}(\epsilon -eV)\tau_{+}\delta_{s,-1}+
e^{-i\varphi_{0}}g^{21}_{L}(\epsilon +eV)\tau_{-}\delta_{s,1},
\label{Eq_GEE}
\end{eqnarray}
where the superscripts denote the matrix elements in the Nambu space and
$P_{\pm}=(1\pm \tau_z)/2$.

In what follows we shall abbreviate $\hat g_{L}(\epsilon +2meV, \epsilon +2neV)
=(m|\hat g_{L}(\epsilon)|n)$ where the right hand side is obtained from 
eq. (\ref{Eq_GEE}) after replacing $\epsilon \to \epsilon +2meV$, 
$\delta_{0,s}\to\delta_{m,n}$,
$\delta_{-1,s}\to\delta_{n,m-1}$, 
and $\delta_{1,s}\to\delta_{n,m+1}$. 
Then we have
\begin{eqnarray} & &
\hat g_{L}(\epsilon ,\epsilon')=\sum_{n} 
\delta(\epsilon -\epsilon'+2neV)(0|\hat g_{L}(\epsilon)|n),
\label{Eq_gL}
\end{eqnarray}

The matrix $M$ (\ref{Seff2}) may also be
represented in a similar form, that is,
\begin{eqnarray} & &
\hat M(\epsilon ,\epsilon')= \sum_{n}\delta(\epsilon -\epsilon'+n2eV)(0|\hat M(\epsilon)|n),
\label{Eq_Minv} 
\end{eqnarray}
where
\begin{eqnarray} & &
 (m| \hat M(\epsilon)|n) = \delta_{m,n} [\epsilon +m2eV+\gamma_{-} 
\nonumber \\
&& -\tau_{z}\tilde{\epsilon}-\frac{ i\Gamma_R }{2}
\tau_{z}\hat g_{R}(\epsilon+m2eV)]- \nonumber \\
&& \frac { i\tau_{z}\Gamma_L}{2} (m|\hat g_{L}(\epsilon )|n).
\label{Eq_Minv1}
\end{eqnarray}

The integration over energy variables 
in the self-consistent equation for $\gamma_{-}$ and in 
the expression for the time averaged current 
is conveniently performed by dividing the whole energy domain into slices 
of width $2eV$ and performing energy integration on an interval
$[0<E<2eV]$. Thus we can use the 
discrete representation (\ref{Eq_Minv1}) and write
\begin{eqnarray} &&
\gamma_{-}=i\frac{U}{2}\int_{0}^{2eV}
\frac{d\epsilon}{2\pi}\sum_{n}{\rm Tr}(n|(\hat M^R)^{-1}\hat M^K(\hat M^A)^{-1}|n), \nonumber \\
&& I=\frac{e\Gamma_L\Gamma_R}{8}\int_{0}^{2eV}
\frac{d\epsilon}{2\pi}\sum_{n}{\rm Tr} 
(n| [(\hat N_{L}
\hat g_{R}^{R} \nonumber \\
&& -(L\leftrightarrow R)])+{\rm h.c.}]|n) .
\label{Eq_gammaJ}
\end{eqnarray}

Let us also note that in the case of $SAN$ junctions the expressions for
the current and for $\gamma_-$ can be simplified further. E.g. it is easy
to observe that in this case eq. (\ref{Eq_Jfinal}) takes the form
\begin{eqnarray}&&
I=\frac{e\Gamma_L\Gamma_R}{2}\int_{-\infty}^{\infty}
\frac{d\epsilon }{2\pi} {\rm Tr}
[((\hat M^{R}(\epsilon))^{-1}\hat f(\epsilon ,V)(\hat
M^{A}(\epsilon))^{-1}
\hat g_{R}^{R}(\epsilon)-  \nonumber \\
& &(\hat M^{R}(\epsilon))^{-1} \hat f(\epsilon ,0)\tau_{z}\hat
g_{R}^{R}(\epsilon))(\hat M^{A}(\epsilon))^{-1}\tau_{z})+{\rm h.c.}],
\label{Eq_JNS}
\end{eqnarray}
where the matrix $\hat f$ has the standard form 
\begin{eqnarray}
\hat f(\epsilon ,V)=\left ( \begin{array}{cc}
\tanh \left(\frac{\epsilon +eV}{2T}\right)& 0\\
 0 & \tanh \left(\frac{\epsilon -eV}{2T}\right) 
\end{array}\right), 
\label{f}
\end{eqnarray} 

Eq. (\ref{Eq_JNS}) can be straightforwardly evaluated since the (Fourier transformed) matrices $(\hat M^{R,A})^{-1}$ depend now only 
on one energy $\epsilon$ ($\hat g_L$ in (\ref{Seff2}) is proportional to 
$\delta (\epsilon -\epsilon')$ in this case) and, hence, can easily be inverted analytically. Similar simplifications can also be performed in the self-consistency eq. (\ref{samosogl}).

\section{Results and Discussion}

\subsection{The $SAN$ junction}

\subsubsection{$s$-wave superconductors}
We start from calculating the differential conductance of an $SAN$
contact assuming the $s$-wave pairing symmetry in a superconducting electrode.
As it was already pointed out above, eqs. (\ref{Eq_JNS}), (\ref{f}) allow
one to proceed analytically. From these equations one obtains the expression 
for current which consists of two parts. The first part originates from the integration over sub-gap energies $\epsilon < \Delta$ and yields the dominating contribution to the current at low
temperatures. The other part comes from integration over energies 
$\epsilon>\Delta$. At low voltages and temperatures (lower than
the gap $\Delta$) this second
part gives a negligible contribution to the current. Considering below the 
sub-gap contribution only, we find
\begin{eqnarray} & & 
I=\frac{e\Gamma_L\Gamma_R}{4}\int_{-\infty}^{\infty}
\frac{d\epsilon }{2\pi} 
{\cal B}(\epsilon) \nonumber \\
&& \left[\tanh \left(\frac{\epsilon +eV}{2T}\right)-
\tanh \left(\frac{\epsilon -eV}{2T}\right)\right],
\label{Ins}
\end{eqnarray}
where at sub-gap voltages and energies one has
\begin{eqnarray} & &  
{\cal B}(\epsilon )= 
 \frac{\Delta^2\theta (|\Delta|-|\epsilon |)}{\Delta^2-\epsilon^2} \nonumber \\
&& \frac{\Gamma_L\Gamma_R}
{\left(\tilde \epsilon^2+\frac{\Gamma_L^2}{4}+
\frac{\Gamma_R^2}{4}\frac{\Delta^2}
{\sqrt{\Delta^2-\epsilon^2}}-\chi \right)^2+
\Gamma_L^2\chi}
\label{B}
\end{eqnarray}
and 
\begin{eqnarray} & &
\chi =
\left(\epsilon +\gamma_{-}+ \frac{\Gamma_R}{2}
\frac{\epsilon}{\sqrt{\Delta^2-\epsilon^2}}\right)^2.
\label{chi}
\end{eqnarray}
In the limit $eV \ll \Delta$ and $T \to 0$ for the conductance 
$G\equiv I/V$ we obtain
\begin{eqnarray} & &
G=\frac{e^2}{h}\frac{\Gamma_L^2\Gamma_R^2}
{\left( \frac{\Gamma_L^2+\Gamma_R^2}{4}+\tilde \epsilon^2 -\gamma_-^2\right)^2+
\gamma_-^2\frac{\Gamma^2_L}{4}}.
\label{linearNS}
\end{eqnarray}
In order to recover the expression for $G$ in the
noninteracting limit in eq. (\ref{linearNS}) one should simply put $\gamma_-=0$.
In a symmetric case $\Gamma_L=\Gamma_R$ and for $\tilde \epsilon \to 0$
eq. (\ref{linearNS}) reduces to the well known result \cite{BTK2}
\begin{eqnarray} & &
G_{NS}=2G_{NN}=\frac{4e^2}{h}.
\label{BTK2}
\end{eqnarray}

In the presence of Coulomb interaction the parameter $\gamma_-$ in
(\ref{B}-\ref{linearNS}) should be determined from the self-consistency
equation (\ref{samosogl}). This equation has a solution provided the
interaction $U$ is not very large, i.e. outside the Kondo regime. 
In the limit of large $U$ the solution of eq. (\ref{samosogl}) is absent, 
which indicates the failure of the MF approximation in the Kondo regime.

Here eq. (\ref{samosogl}) was solved numerically for a given
set of the system parameters. In our numerical analysis we chose
$\Gamma_L=\Gamma_R=\Gamma=0.35\Delta$ and considered the most interesting
sub-gap voltage bias regime $eV \lesssim 2\Delta$ in the low
temperature limit $T \to 0$.  The values of the Hubbard repulsion parameter
$U$ were fixed to be $U=2.45\Delta$ and $U=2.72\Delta$. 
For convenience we scaled our equations expressing the parameters  
$\epsilon$, $U$, $\Gamma_{L,R}$ and $T$ in units of $\Delta$. After that the
current and the conductance are normalized respectively in units of $\Delta e
/\hbar$ and $e^2/2 h$. 

\begin{figure}[htb]
\includegraphics[width=0.45\textwidth,keepaspectratio]{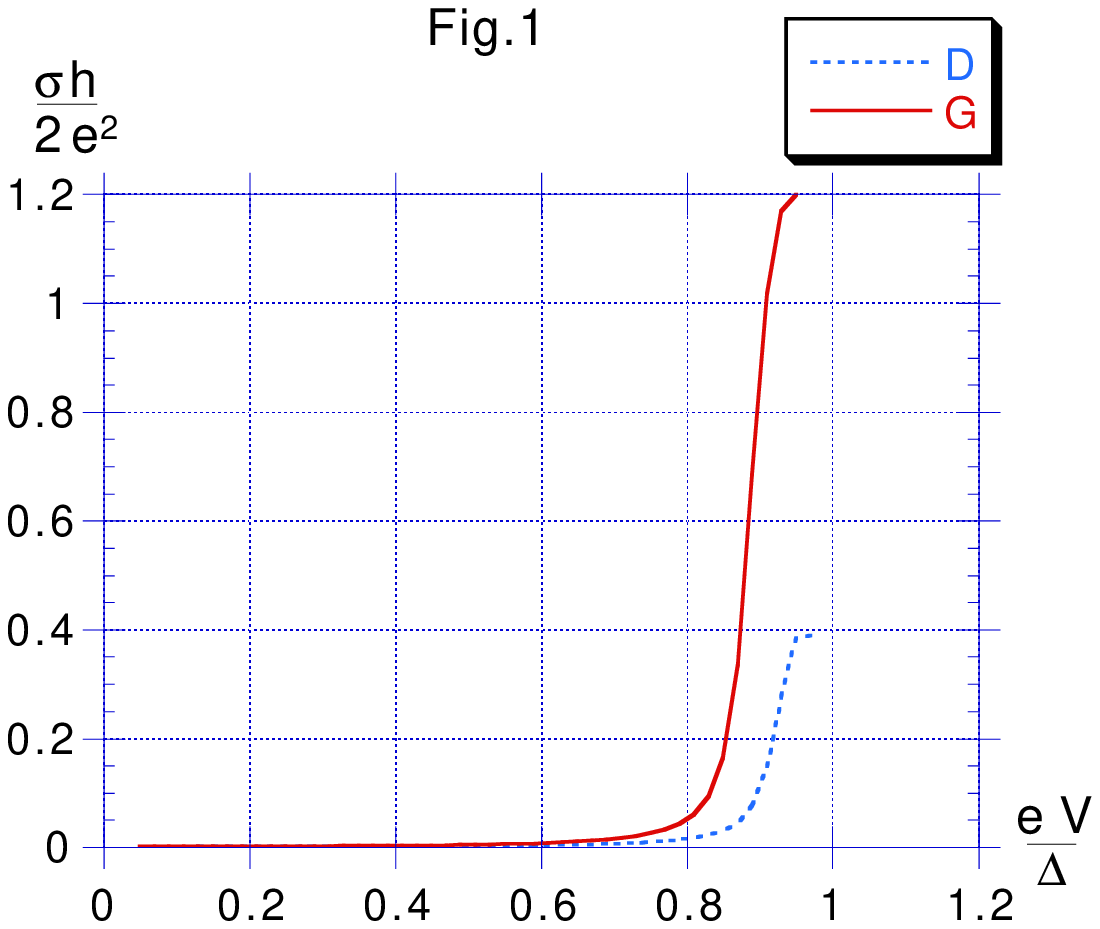}
\caption{Differential conductance of an $SAN$
junction with $s$-wave
symmetry superconductor. The figure displays the
dependence of Andreev conductance on
the applied voltage (in units of superconducting pair potential)
for $U=2.450$ (solid line curve G)
and $U=2.713$ (dotted line curve D). The barrier transparency
is $\Gamma=0.35$ and the dot level energy is
$\epsilon_{0}=-1.5$. }
\label{fig1}
A similar study of an interplay between two-electron tunneling
and Coulomb effects in $SIS$ junctions was carried out in Ref. \onlinecite{Jens}. 

There exists also a certain analogy between our results and those
obtained for superconductor-ferromagnet ($SF$) junctions \cite{JB}.
Here the repulsion parameter $U$ plays a role 
similar to that of an exchange term in $SF$ systems: in both cases 
the sub-gap conductance can be tuned by changing this parameter
in a way that a smaller value of $U$ corresponds
to a weaker exchange field. In contrast to our system, however, changing of
the exchange field in $SF$ junctions leads to smooth variations of
the sub-gap conductance \cite{JB}.

Let us also note that here we 
do not consider the Kondo limit \cite{gol2,fazio,ambe}, 
in which case a zero bias conductance peak is expected. 
This peak appears, simply, because in 
the Kondo limit and at $V=0$ there exists an 
open channel between the normal metal and the 
superconductor. Here the Kondo temperature
$T_{K}$  is assumed to be small and,
hence, the zero bias peak is absent.

Let us now briefly consider the limit of large bias voltages $eV \gg \Delta$.
In this case the current may be represented as a sum of two terms $I=I_1+I_2$. The
term $I_1$ is determined by the expression similar to (\ref{Ins}) which now includes 
the contribution from energies above the gap. We find  
\begin{eqnarray} & & 
I_1= \nonumber \\
&& \frac{e\Gamma_L\Gamma_R}{4}\int_{-\infty}^{\infty}
\frac{d\epsilon }{2\pi} 
({\cal B}(\epsilon)+{\cal B}_1(\epsilon)) \nonumber \\
&& \left[\tanh \left(\frac{\epsilon +eV}{2T}\right)-
\tanh \left(\frac{\epsilon -eV}{2T}\right)\right].
\label{Ins1}
\end{eqnarray}
Here ${\cal B}(\epsilon)$ is again given by eq. (\ref{B})
while the function 
${\cal B}_1(\epsilon)$ reads
\begin{eqnarray}&&
{\cal B}_{1}(\epsilon )= 
\frac{2|\epsilon|\theta 
(|\epsilon|-|\Delta|) }{\epsilon^2-\Delta^2} \nonumber \\
&& \frac{[\tilde \epsilon^2+(\epsilon +\gamma_{-})^{2}+\chi_{1}]}
{\left(\tilde \epsilon^2-(\epsilon +\gamma_{-})^{2}+\chi_{1}
 \right)^2+
(\epsilon +\gamma_{-})^{2}(\Gamma_L +\Gamma_R 
\frac{|\epsilon|}{\sqrt{\epsilon^2-\Delta^2} })^2 }.
\label{B1}
\end{eqnarray}
Here we also defined 
\begin{eqnarray} & &
\chi_{1} =
\frac{1}{4}(\Gamma^{2}_{L}+\Gamma^{2}_{R} + 2\Gamma_L \Gamma_R
\frac{|\epsilon|}{\sqrt{\epsilon^2-\Delta^2}}).
\label{chi1}
\end{eqnarray}
The other contribution $I_2$ is proportional 
to the level position $\tilde{\epsilon}$.
One obtains
\begin{eqnarray} & &
I_{2}=\frac{e\Gamma_L\Gamma_R}{4}\int_{-\infty}^{\infty}
\frac{d\epsilon }{2\pi} \nonumber \\
&& {\cal B}_{2}(\epsilon)[\tanh \left(\frac{\epsilon +eV}{2T}\right)+
\tanh (\frac{\epsilon -eV}{2T})-2\tanh (\frac{\epsilon }{2T})].
\label{Ins2}
\end{eqnarray}
The expression for ${\cal B}_2 (\epsilon)$ can be obtained from eq. (\ref{B1}) if 
one replaces the term in the square brackets by the expression 
$-2\tilde{\epsilon}(\epsilon +\gamma_{-})$. 

The above results together with the self-consistency equation for $\gamma_-$ 
provide a complete description for the $I-V$ curve of an $SAN$ junction in the
presence of interactions. In all interesting limits the energy integrals
in (\ref{Ins1}), (\ref{Ins2}) can be carried out and the corresponding expressions
for the current can be obtained. These general expressions, however, turn out to
be quite complicated and will not be analyzed in details further below.

Here we just demonstrate that in the noninteracting limit $\gamma_-=0$ our 
results reduce to the results already well known in the literature.
In the
leading order approximation eqs. (\ref{Ins1}), (\ref{B1}) yield the
standard Breit-Wigner formula
\begin{eqnarray} & &
\sigma=\frac{2e^2}{h}\frac{\Gamma_L\Gamma_R}
{\frac{(\Gamma_L+\Gamma_R)^2}{4}+\tilde \epsilon^2 }.
\label{linearNN}
\end{eqnarray}
After setting  $\tilde{\epsilon}=0$ and 
$\Gamma_L=\Gamma_R \gg \Delta$ in eqs. (\ref{Ins1}),
(\ref{B}), (\ref{B1}) in the limit
$eV \gg \Delta$ one easily obtains the contributions to the current equal to 
$2G_{NN}\Delta/e$ and  $G_{NN}(V-2\Delta/3e)$ 
respectively from the sub-gap energies
(${\cal B}$) and from energies above the gap (${\cal B}_1$). The sum of these
contributions yields the standard result
\begin{eqnarray} & &
I=G_{NN}(V +4\Delta/3e).
\label{sh}
\end{eqnarray}
The second term  represents the so-called excess current which originates from the 
mechanism of Andreev reflection. It follows from our general analysis that 
in quantum dots this current is also affected by Coulomb interaction.
\subsubsection{Superconductors with unconventional pairing}
Since the order parameter $\Delta$  for $p$- 
and $d$-wave superconductors
is not isotropic, the magnitude of the current is sensitive 
to the junction geometry. 
As discussed before, here we consider the system of two
planar superconducting (or normal) strips with electron
tunneling between them along the $x$ axis through the 
dot located at $x=y=0$. 
For $d$-wave superconductors we 
choose the nodal line of the pair potential 
on the Fermi surface to coincide with the 
tunneling direction (figure \ref{fig2}), such that 
$\Delta =v_{\Delta}p_{p_{F}}\sin2\alpha $.
\begin{figure}[htb]
\centering
\includegraphics[
height=0.2\textheight,
keepaspectratio]{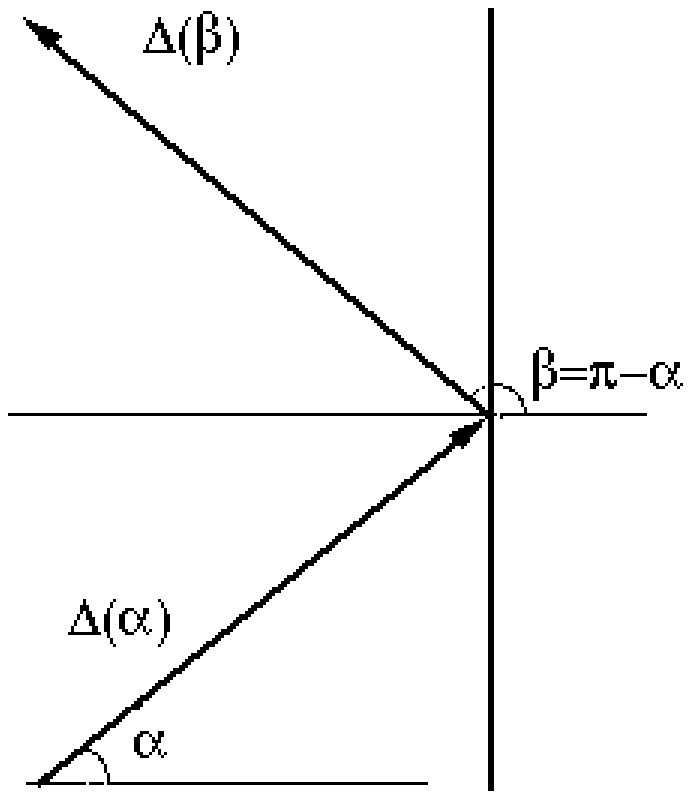}
\caption{Schematic geometry of the junction. Incoming and reflected
electron-like excitations are moving in an angle-dependent pair potential
which can have different signs for these quasiparticles.}
\label{fig2}
\end{figure}
The direction of tunneling  
corresponds to the angle $\alpha=0$. 
For spin-triplet superconducting 
states the order parameter is an odd 
vector function of momentum and a $2\times 2$ 
matrix in spin space. We choose to 
represent it by a time reversal 
symmetry breaking state \cite{sig} 
which is off-diagonal in spin indices. 
In the geometry of figure \ref{fig2}, $\alpha$ is the azimuthal angle
in the $x-y$ plane and the order parameter can
approximately be represented 
as $\Delta=\Delta_{0}\exp(i\alpha)$. 
This order parameter can possibly describe pairing in a
superconductor $Sr_{2}RuO_{4}$ which was recently discovered \cite{maeno}. 
The pair potential so chosen within the geometry of the 
junction may have different signs
for incoming and reflected quasiparticles moving at the  
angles $\alpha$ and $\pi-\alpha$, 
respectively. This fact significantly affects the
scattering process and causes the 
formation of a zero energy (mid-gap) 
bound state \cite{Hu} centered at the boundary. For 
this state we calculate the Green 
function $\hat G $ which, 
like in the case of $s$-superconductors, satisfies
eq. (\ref{Eq_In}) with a $\delta$-function 
on the right side and require $\hat G$ to vanish
at $x=0$. The distinction of solutions 
for $d$- or $p$-wave superconductors from
those found above for the $s$-wave case is 
due to the sign change of the pair potential: 
reflected quasiparticles propagate 
in a pair potential of an opposite sign as compared to
$\Delta$ ``seen'' by incoming quasiparticles. 
The equilibrium retarded and advanced Eilenberger functions $\hat g^{R,A}$ 
for $p$-wave superconductors read
\begin{eqnarray}
\hat g^{R,A}(\epsilon)=\frac{\sqrt{(\epsilon \pm i0)^2-\Delta^2}-\tau_{+}\Delta+\tau_{-}\Delta^*}{\epsilon \pm i0}.
\label{Eq_gpwave}
\end{eqnarray}

The $I-V$ curves for $SAN$ junctions in the case of 
$p$-wave superconductors are  
remarkably distinct from those found for the $s$-wave case 
(cf. Figs. 1 and 3). This difference is predominantly 
due to the surface bound surface which exists in the $p$-wave case
and causes the conductance peak in the sub-gap region. 
Due to electron-electron 
repulsion this peak is split and
appears at $V\neq 0$, see figure \ref{fig3}.
\includegraphics[width=0.45\textwidth,keepaspectratio]{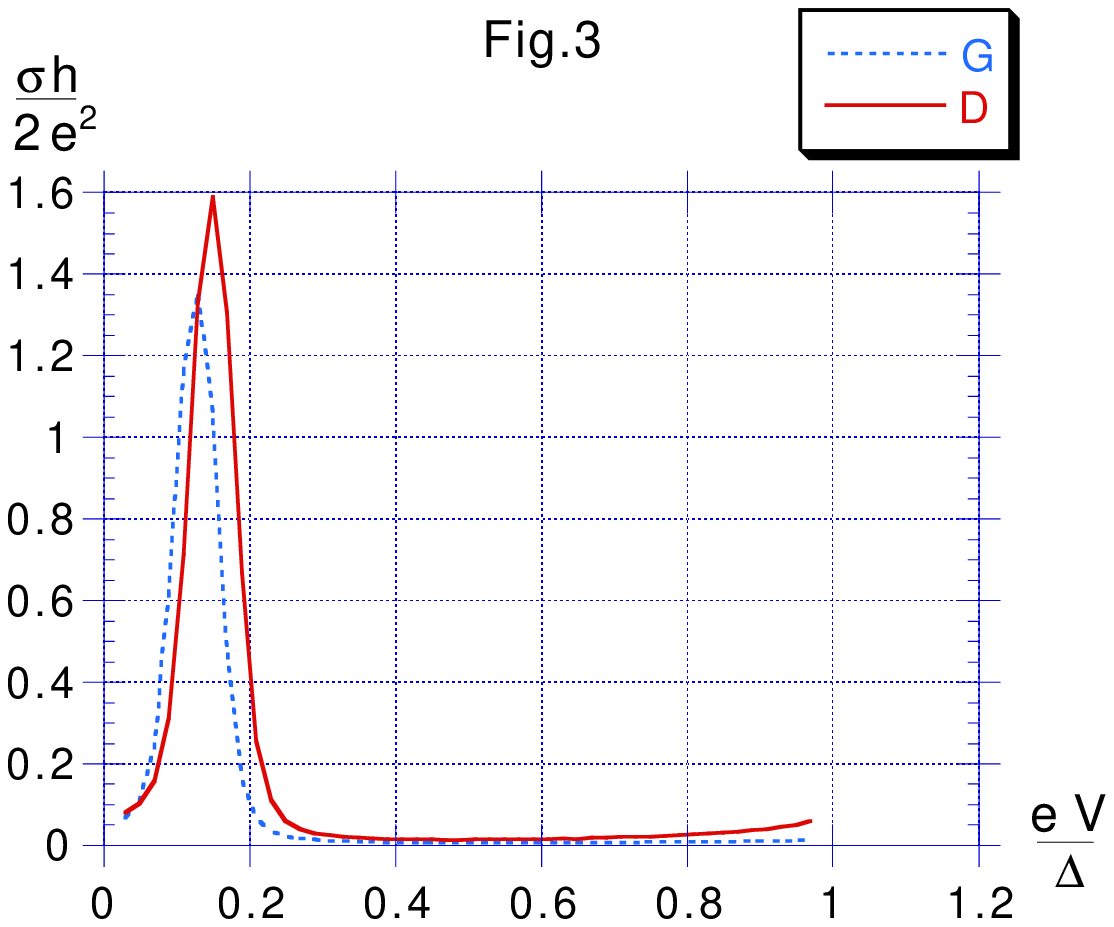}
\caption{Same as in figure \ref{fig1} but for a $p$-wave
symmetry superconductor.
The parameters $U$, $\Gamma$ and
$\epsilon_{0}$ are identical to those of
figure \ref{fig1}.}
\label{fig3}
\end{figure}

For the above parameters the value $\gamma_-$ was found to be 
$\gamma_- \approx 1.2\Delta$
with variations (depending on the voltage) within $10 \div 15$ per cents.
Thus within a reasonable accuracy in eqs. (\ref{B}-\ref{linearNS}) one can
consider $\gamma_-$ as constant. On the other hand, even though the variations 
of $\gamma_-$ with voltage are not
large in magnitude, sometimes they occur over a small voltage interval.
Therefore such variations may have a considerable impact on the
differential conductance $\sigma =dI/dV$ and should also be taken into
account. This was done within our numerical analysis. The corresponding 
results are presented in figure \ref{fig1}.  

We observe that for given parameters the conductance virtually vanishes 
in the substantial part of the sub-gap region. Note, however, that
at voltages close to but still {\it smaller} than $\Delta /e$
the differential conductance $\sigma$ increases sharply. This feature 
can easily
be understood as a result of interplay between Coulomb blockade and two electron
tunneling effects. It is well known \cite{BTK2} that the sub-gap conductance
in $SN$ junctions is caused by the mechanism of Andreev reflection 
during which the charge $2e$ is transferred between the electrodes. 
Without interaction eq. (\ref{BTK2}) holds down to $V \to 0$ while 
in the presence of interaction and at $T=0$ two electron tunneling 
process is completely blocked for $eV \leq 2E_C$ (see e.g. 
\cite{Hek,HekNaz,Z94}), 
where $E_C=e^2/2C$ and $C$ is the characteristic junction capacitance.
For larger voltages $eV > 2E_C$ two electron tunneling cannot anymore be
blocked by Coulomb effects and the current can flow through the system.
Obviously, for $2E_C < \Delta$ this results in a finite sub-gap
conductance at voltages $2E_C<eV<\Delta$. A similar behavior is
obtained here (see figure \ref{fig1}) with an important difference, however, that
within our model the characteristic Coulomb energy $E_C$ is
obtained self-consistently and, 
hence, it depends not only on the interaction
parameter $U$ but also on the voltage $V$ as well as on $\Gamma$ and
$\epsilon_0$. In the case of metallic $SIN$
tunnel junctions the effect 
of Coulomb interaction on the $I-V$ curve was studied
in details both theoretically \cite{Z94} 
and experimentally \cite{Tinkham}.
Here again, the repulsion attenuates the conductance,
which is larger for $U=2.45\Delta$ than for $U=2.72\Delta$.

\subsection{The $SAS$ junction}

Let us start from a noninteracting case $U=0$ and briefly consider 
pure resonant tunneling at the Fermi level, i.e. set $ \epsilon_0 \to 0$. This 
situation corresponds to a ballistic $SNS$ junction with only one
conducting channel. The $I-V$ curves of ballistic
$SNS$ junctions were intensively studied in the past  
\cite{arnold,zai,Z83,gun,averin,bez,cuevas,LY,shumeiko}. If the relevant 
energies are
small as compared to $\Gamma$ 
(for short junctions this condition usually means $\Gamma > \Delta$),
$S_{\rm dot}$ in (\ref{Jeff}) can be dropped and one gets 
$\langle \bar{c}c\rangle=\hat g_+^{-1}\tau_z/\Gamma $. 
Eq. (\ref{tok}) then yields,
\begin{eqnarray} & &
I= \frac{e}{2}{\rm Tr}\tau_z\hat g_-\hat g_+^{-1}|_K .
\label{zait}
\end{eqnarray} 
Note that here the tunneling rate $\Gamma $ just cancels out.
In the many channel limit eq. (\ref{zait}) coincides with the 
quasiclassical result \cite{zai,Z83}. 
For a constant bias $V$ the matrix $\hat g_+^{-1}$ can be evaluated
analytically \cite{gun}, yielding the $I-V$ curve of a
ballistic SNS junction. In particular, in the zero bias limit $V \to 0$ 
and for $\Gamma \gg \Delta$
one recovers the MAR current \cite{gun}: 
\begin{eqnarray} & &
I_{AR}=\frac{2e^2}{h}\frac{2\Delta}{eV}V=\frac{4e\Delta}{h}.
\label{Iar}
\end{eqnarray} 
The corresponding explicit calculation performed within our
formalism is presented in Appendix. We also note that in the limit of small 
tunneling rates $\Gamma_{L,R} < \Delta$ resonant effects gain importance.
An interplay between MAR and resonant tunneling in the absence of
Coulomb interaction was studied in Refs. \onlinecite{LY,shumeiko}.

Let us now turn to $SAS$ junctions with interactions.

\subsubsection{$s$-wave superconductors}
In order to calculate the  
sub-gap current in the case of an $SAS$ junction 
one has first to find the solution of the self-consistency 
equations (\ref{Eq_gammaJ}). This requires the
inversion of the  matrix $\hat M$ 
in the energy and spin spaces. If the number of modes 
for each energy in the interval  $[0,2eV]$
is cut off at some integer $m$, the size of the pertinent matrices 
is  $(4m+2)\times (4m+2)$. 
The number $m$ of the energy slices 
has to be adjusted in such a way 
that the results become insensitive to it. This requires larger $m$
for smaller voltages because quasiparticles can escape the gap
region after undergoing a large number of Andreev reflections.

\begin{figure}[htb]
\includegraphics[width=0.45\textwidth,keepaspectratio]{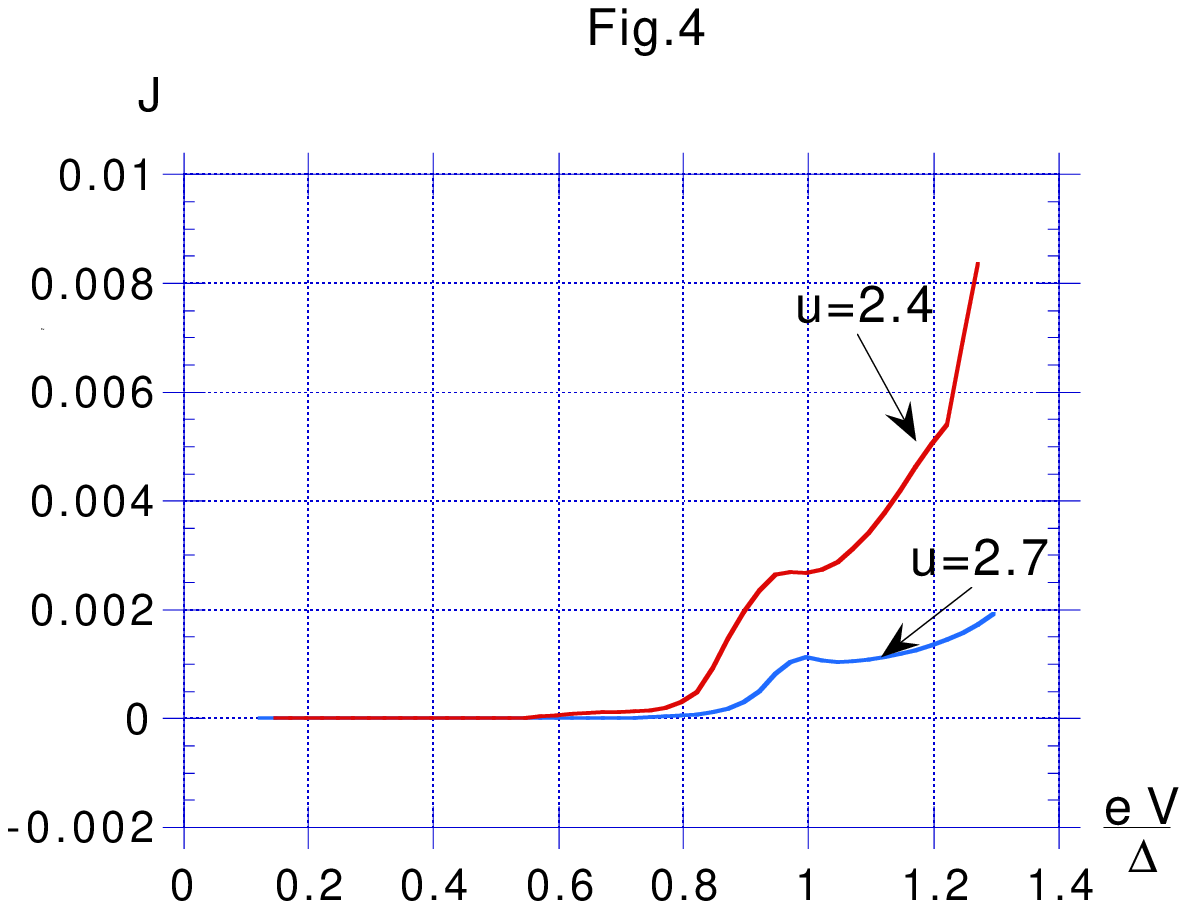}
\caption{The subgap tunneling current versus voltage
for an $SAS$ junction for which the
superconductors pair potential has an $s$-wave symmetry.
The parameters are $U=2.4,2.7$, $\epsilon =-1.5$ and  $\Gamma=0.6$.}
\label{fig4}
\end{figure}

The $I-V$ characteristics for
tunneling between two $s$-wave superconductors is displayed in
figure \ref{fig4}. 
The transparency of the 
junction is chosen to be $\Gamma=0.6\Delta$ and the current is 
evaluated for $U=2.4\Delta$ and $U=2.7\Delta$. One observes that
at relatively low bias voltages $eV \lesssim 0.8\Delta$ for $U=2.4\Delta$
and $eV \lesssim (0.9\div 0.95)\Delta$ for $U=2.7\Delta$ the sub-gap
current is essentially suppressed. For higher voltages 
the sub-gap current increases rather sharply, 
as a result of an interplay between Coulomb blockade and multiple
Andreev reflections. The latter mechanism manifests itself in 
the occurrence of the sub-harmonic peaks 
in the differential conductance. 
Due to interaction, the positions of these peaks are shifted
relative to those in the noninteracting case $eV=2\Delta/n$, where 
$n$ is the number of Andreev reflections. 
As can be seen in figure \ref{fig4}, increasing $U$
results in a larger shift of peak positions. 

The behavior observed in figure \ref{fig4} has a transparent physical interpretation
which can be summarized as follows. It is well known that in the absence of
interactions the sub-gap conductance of superconducting junctions is determined
by the process of multiple Andreev reflections (MAR) with the effective number of
such reflections $n \approx  2\Delta /eV$. For the process with given $n$ the
charge $(n+1)e$ is transferred between the electrodes. The relevant number $n$
is obviously large at small voltages $eV \ll \Delta$. Let us now turn on the
interaction which strength is again characterized by some effective charging
energy $E_C=e^2/2C$. Clearly, in this case MAR cycles with large $n$ will be
suppressed due to Coulomb effects much stronger than, say, one electron
processes simply because in the case $n \gg 1$ the charge much larger 
than $e$ is being
transferred. E.g. at $T=0$ the single electron processes are blocked for $eV
\leq E_C$, while MAR will be fully suppressed already at higher voltages $eV
\leq (n+1)E_C$. This implies that for relatively small voltages the sub-gap
conductance can be effectively suppressed even if the Coulomb energy $E_C$ is
small as compared to $\Delta$. This is exactly what we observe in figure \ref{fig4}.

Let us now recall that the
sub-harmonic peaks on the $I-V$ curve occur at the voltage values for which
quasiparticles participating in MAR cycle with $n$ reflections start leaving
the pair potential well. In the presence of interactions this becomes
possible if the energy $eVn$ gained by a quasiparticle (hole) after $n$
reflections is equal to $2\Delta +(n+1)E_C$. This condition immediately
fixes the sub-harmonic peak positions at
\begin{eqnarray} & &
V_n=\frac{E_C}{e}+\frac{2\Delta +E_C}{en},
\label{peaksEc}
\end{eqnarray}
i.e. due to interaction the sub-harmonic peaks are shifted by $e(1+1/n)/2C$
to larger voltages as compared to the noninteracting case. This feature is 
fully reproduced within
our numerical analysis (see figure \ref{fig4}), which also allows to self-consistently
determine $E_C$ as a function of the parameters $U$, $\Gamma$, $V$ and 
$\epsilon_0$.

Combining the Coulomb blockade condition $eV \leq (n+1)E_C$ with the
effective number of Andreev reflections at a given voltage 
$$
n=\frac{2\Delta +E_C}{eV-E_C}
$$
(here $eV >E_C$, in the opposite case no MAR is possible) one easily
arrives at the condition
\begin{eqnarray} & &
eV \leq eV_{th}=E_C\left[1+\sqrt{1+\frac{2\Delta}{E_C}}\right].
\label{threshold}
\end{eqnarray}
This condition determines the voltage interval within which the sub-gap
conductance is suppressed due to the Coulomb effects. In the limit $E_C \ll
\Delta$ the corresponding voltage threshold is $eV_{th} 
\simeq \sqrt{2\Delta E_C} \gg E_C$. Finally, by setting $eV_{th}\geq 2\Delta$ 
into eq. (\ref{threshold}) one immediately finds that for
$$
E_C \geq \frac{2\Delta}{3}
$$
the sub-gap conductance is totally suppressed due to Coulomb 
interaction and no sub-gap current can flow through the system.    

With the aid of eq. (\ref{threshold}) we can (roughly) 
estimate the effective value of
$E_C$ for the parameters used in our numerical calculations. From the $I-V$
curves presented in figure \ref{fig4} we find $E_C \approx 0.2\Delta$ for $U=2.4\Delta$
and $E_C \approx 0.25\Delta$ for $U=2.7\Delta$. Obviously, these
values are smaller than $2\Delta /3$ and, hence, the sub-gap conductance is
not totally suppressed at intermediate voltages. Substituting the above  
values of $E_C$ into eq. (\ref{peaksEc}) we can also estimate the magnitudes
of the peak shifts with respect to their ``noninteracting'' values. For
$U=2.4\Delta$ the peaks are shifted by $\sim 0.3\Delta$ for $n=2$ and
$\sim 0.26 \Delta$ for $n=3$. Analogous values for $U=2.7\Delta$ are
respectively $\sim 0.38\Delta$ and $\sim 0.33 \Delta$. These values are in 
a reasonably good agreement with our numerical results.

In the limit of high voltages $eV \gg \Delta$ the $I-V$ curves for $SAS$
junctions are analogous to those for $SAN$ ones except the excess current
is two times larger.

\subsubsection{Superconductors with unconventional pairing}
\begin{figure}[htb]
\includegraphics[width=0.45\textwidth,keepaspectratio]{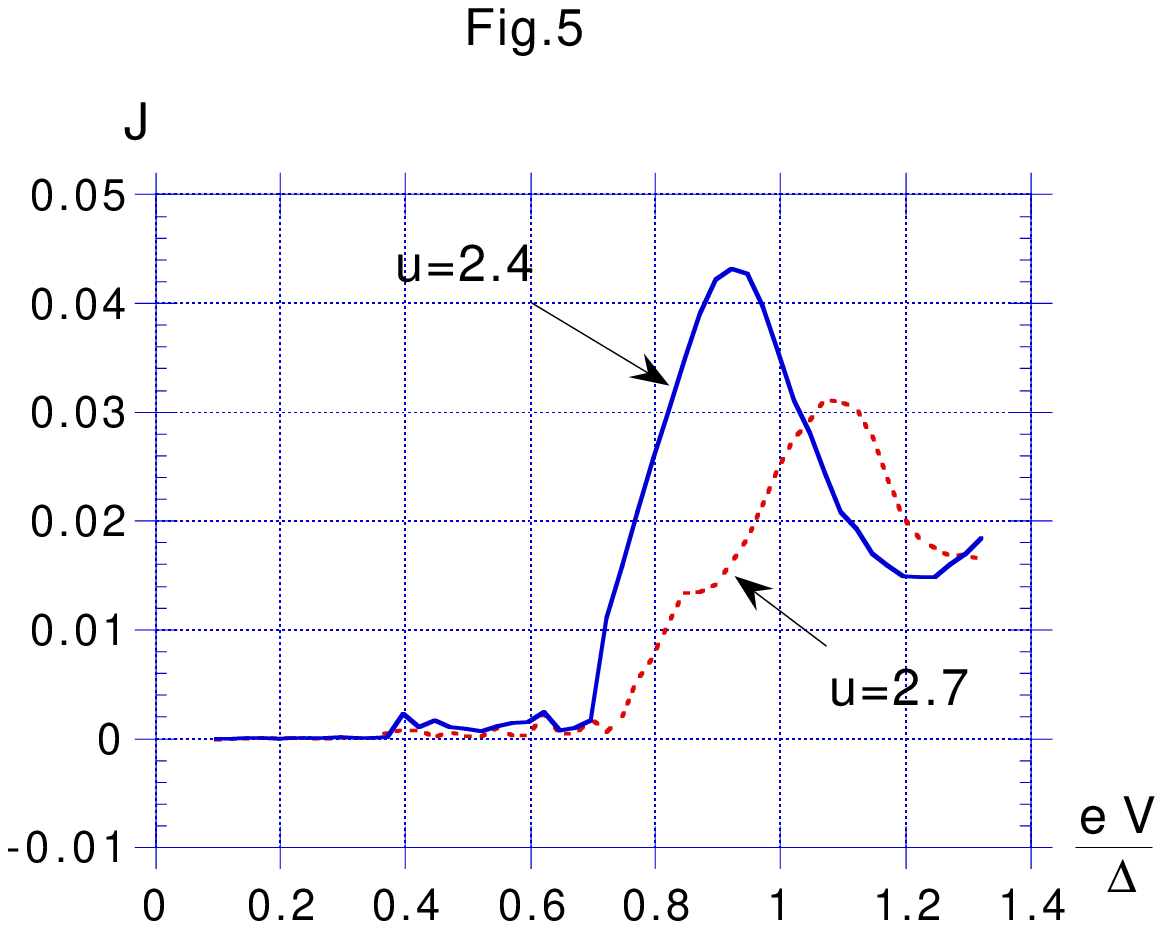}
\caption{Same as figure \ref{fig4} but for
superconductors pair potential with $p$-wave symmetry.
The parameters are $U=2.4$ (dotted curve ), $U=2.7$ (solid curve),
 $ \epsilon =-1.5$ and  $\Gamma=0.6$.}
\label{fig5}
\end{figure}

Similarly to the case of $SAN$ junctions, there exists an 
important difference in the tunneling current between $SAS$ 
junctions depending on whether the order parameter in the
electrodes is of $s$- or $p$-wave symmetry. 
The $I-V$ curve for the latter case is depicted in figure \ref{fig5}.
 We observe that 
the sub-gap current for $p$-wave superconductors
is considerably larger than for $s$-wave ones, roughly by  
$I_{max}^{(p)}/I_{max}^{(s)} \approx 8$. 
On the other hand, the effect of the Coulomb repulsion $U$
is rather similar. For $U=2.7\Delta$ 
the current is suppressed compared to its value at $U=2.4\Delta$.
Beside the distinction of magnitudes, there is
a novel additional structure in the $I-V$
curves for $p$-wave superconductors 
which is related to the presence of a surface bound state. 
Comparing the results presented in Figs. 4 and 5 
we observe that in the latter case,
the current peaks at a certain bias voltage. 
This implies a negative differential
conductance, which is the hallmark of
resonant tunneling (contributed by the bound state). 

Our analysis of the junctions formed by $p$-wave superconductors 
can be straightforwardly extended to  
the case of $d$-wave pairing. The $I-V$ curves and the sub-harmonic
gap structure in junctions with $d$-wave superconductors in the
absence of Coulomb interaction was recently studied 
(see e.g. Ref. \onlinecite{wendin} and other Refs. therein). 
Near zero bias the $I-V$ curves \cite{wendin} exhibit a 
current peak (equivalently, negative differential conductance) related 
to the presence of the mid-gap surface states. 
Notice, that in such systems the symmetry restricts the current, so 
that the contribution from the bound mid-gap states may vanish if, for instance,
one assumes ${\cal T}^2_{L,R}$ to be independent of $v_x$. 
As we have already argued before (see also Ref. \onlinecite{zaikin}),
it might be essential to take the dependence of tunneling matrix elements
on $v_{x}$ into account already for point contacts. One can also
consider the impurity model different from a point-like defect. Such a situation 
can be realized e.g. by artificially-induced defects \cite{yazdani}. 
The spectroscopy of
$Bi_{2}Sr_{2}CaCu_{2}O_{8}$ surfaces indicates that such defects
appear to be more extended in STM imaging. In this case one can expect
non-zero contribution from mid-gap level also in $d$-waves superconductors. 
Here, again, the electron-electron repulsion 
shifts the peak positions from their ``noninteracting'' values  $eV=2\Delta/n$
to higher voltages.
It is quite likely that this interaction-induced shift was observed in the 
experiment \cite{brag}. 

\section{Conclusions}
In this paper the 
tunneling between two superconductors 
or between a superconductor and normal 
metal through an Anderson-type quantum dot is investigated.
Special attention is devoted to analyze
the implications of the Coulomb repulsion between 
electrons in the dot on the tunneling process. 
The Andreev conductance for an $SAN$ junction 
and the sub-gap current in an $SAS$ junction 
are calculated and elaborated upon. 
The theoretical treatment requires a combination 
of the Keldysh non-equilibrium Green function 
and path integral formalism and the dynamical mean field approximation. 
We derive general expressions for the effective action and
the transport current through the system. 
These expressions are then employed in order to obtain 
a workable formula for the current.  
The latter is then calculated analytically and numerically 
for a certain set of energy parameters.  

The main results of the present research 
can be summarized as follows: 1) When one
of the electrodes is a normal metal 
(an $SAN$ junction) 
the gap symmetry structure is exhibited
in the Andreev conductance. For $p$-wave superconductors, it shows 
a remarkable peak for voltages in the 
sub-gap region. For $s$-wave superconductors, on the 
other hand, the position of the peak
is shifted towards the gap edge. It is further demonstrated
that if the Hubbard repulsive 
interaction increases the current
peak at the gap edge is strongly suppressed.
2) The dynamics of tunneling between 
two superconductors (an $SAS$ junction) is more
complicated. 
For $s$-wave superconductors the usual 
peaks in the conductance that originate from multiple
Andreev reflections \cite{arnold} 
are shifted by interaction to higher values of $V$. 
They also suffer sizable suppression as the 
Hubbard interaction strength increases. The sub-gap current in this
case may describe the low energy channels in break junctions \cite{scheer}. 
For $p$-wave superconductors, the sub-gap
current is much larger than in the $s$-wave case 
and the $I-V$ characteristics 
exhibits a novel feature: the occurrence of 
mid-gap bound state results in a peak in the current, 
that is, a negative differential conductance.

\noindent
{\bf Acknowledgments}: This research is supported in part by grants from the
Israeli Science Foundation ({\em Center of Excellence and
Non-Linear Tunneling}), the German-Israeli
DIP foundation {\em Quantum Electronics in Low Dimensional 
Systems} and the US-Israel BSF grant {\em Dynamical Instabilities 
in Quantum Dots}.

\section {Appendix}
Below we will derive the result (\ref{Iar}) within the framework of the
formalism developed in the present paper. Consider a quantum dot between
two $s$-wave superconductors and assume that the interaction is negligibly
small $U \to 0$. For the sake of simplicity we will also set
$\Gamma_L=\Gamma_R=\Gamma$. The result (\ref{Eq_Jfinal})  can be
expressed as a sum of two terms $I=I_{AR}+I_{qp}$, where 
\begin{eqnarray}&&
I_{AR}=-\frac{e\Gamma^{2}}{4h}\int_{0}^{2eV}d\epsilon 
{\rm Tr}[(\tilde N_{R})^{12}(\tilde g_{L}^{A})^{21} \nonumber \\
&&-(\tilde N_{L})^{12}(\tilde g_{R}^{A})^{21} - 
(\tilde N_{R})^{21}(\tilde g_{L}^{R})^{12}+
(\tilde N_{L})^{21}(\tilde g_{R}^{R})^{12}],\\ 
&& I_{qp} =
-\frac{e\Gamma^{2}}{4h} \int_{0}^{2eV}d\epsilon 
{\rm Tr}[(\tilde N_{R})^{11}(\tilde g_{L}^{A}-g_{L})^{R})^{11} \nonumber \\
&& -(\tilde N_{L})^{11}(\tilde g_{R}^{A}-\tilde g_{R})^{R})^{11}] . 
\end{eqnarray}
Here $I_{AR}$ is the sub-gap (Andreev reflection) contribution 
to the averaged current while $I_{qp}$ is defined by the excitations above  
the gap. In (A.1)-(A.2) we defined the Green-Keldysh matrices $\tilde
g=i\tau_{z}\hat g $ with     
 \begin{eqnarray} &&
\tilde{g}_{R,L}^{R,A}(\epsilon)=F^{R,A}(\epsilon) 
(\epsilon+\tau_{+}\Delta+\tau_{-}\Delta^*),\\ 
&& F^{R,A}(\epsilon)=\frac{\theta (|\Delta|-
|\epsilon|)}{\sqrt{\Delta^2-\epsilon^2}}\pm 
i{\rm sign}(\epsilon)\frac{\theta 
(|\epsilon|-|\Delta|)}{\sqrt{\epsilon^2- 
\Delta^2}},\\ 
&& \tilde{g}_{R,L}^{K}(\epsilon)=(\tilde{g}_{R,L}^{R}(\epsilon)- 
\tilde{g}_{R,L}^{A}(\epsilon))\tanh(\epsilon /2T). 
\label{Eq_gs} 
\end{eqnarray}  
We also defined 
\begin{eqnarray} & &
\tilde N_{L,R}^{ij}=(\tilde M^{R}\tilde g^{K}_{L,R}\tilde
M^{A})^{i,j}, \;\;\;\;\; \tilde M^{R,A} \equiv (M^{R,A})^{-1},
\end{eqnarray}
where the superscripts stand for the spin indices in Nambu space and Tr
denotes the remaining trace over (discrete) energies which are scaled to
$\Delta$ throughout this Appendix.
 
Consider the limit of small voltages $eV \ll \Delta$. In this limit the sub-gap
current $I_{AR}$ can be rewritten in the form 
\begin{eqnarray} &&
I_{AR}=-\frac{e^2V}{h} \nonumber \\
&& \sum_{m,n}[\tilde g^{K}(E_{m})((m|(\tilde
M^{A})^{i2})|n) \nonumber \\
&& (n+1|(\tilde M^{R})^{1i}|m)F^{A}(E_{n}^{+}) \nonumber \\
&& -(m|(\tilde M^{A})^{i1}|n)(n-1|
(\tilde M^{R})^{2i}|m)F^{R}(E_{n}^{-}))
\nonumber \\
&& \tilde g^{K}(E_{m}^{-})((m|(\tilde
M_{A})^{12})|n) \nonumber \\
&& (n|(\tilde M^{R})^{11}|m)F^{A}(E_{n})-
(m|(\tilde M^{A})^{11}|n) \nonumber \\
&& (n|(\tilde M^{R})^{21}|m)F^{R}(E_{n}))
\nonumber \\
&& \tilde g^{K}(E_{m}^{+})((m|(\tilde
M^{A})^{22}|n) \nonumber \\
&& (n|(\tilde M^{R})^{12}|m)F^{A}(E_{n})-
(m|(\tilde M^{A})^{21}|n) \nonumber \\
&& (n|(\tilde M^{R})^{22}|m)F^{R}(E_{n}))].
\end{eqnarray}
Here we denote $E_{n}^{\pm}=eV(2n\pm 1)$ and $E_{n}=2eVn$. 
We also included  $\Gamma/2$ 
into the definition of $\tilde M$ and omitted terms non-diagonal in the spin
indices because these terms are small in the limit $eV \ll \Delta$. At $T \to
0$ the summation over $m$ is reduced to just one term with the maximum number 
$m_{0}$ determined by the condition: $|E_{m_{0}}|= 1$.

It is straightforward to evaluate the matrices $(m_{0}|\tilde M^{i,j}|n)$ for
sufficiently large $\Gamma > \Delta$ and $\epsilon_{0} \to 0$.  In this case
$\tilde M^{i,j}$ satisfy the following approximate equations  
\begin{eqnarray}&&
(m|(\tilde M^{R})^{11}|m_{0})(E_{m}^2/4-2)= \nonumber \\
&& -E_{m}\delta_{m,m_{0}}/2F^{R}(E_{m}) \nonumber \\
&&+(m-1|(\tilde M^{R})^{11}|m_{0})+(m+1|(\tilde M^{R})^{11}|m_{0}),\\
&& (m|(\tilde
M^{R})^{12}|m_{0})=-4(m|(\bar{\tilde M}^{R})^{12}|m_{0}) \nonumber \\
&& (E_{m}F^{R}(E_{m})
E_{m_{0}}^{-}F^{R}(E_{m_{0}}^{0})^{-1},\\
&& (m|(\bar{\tilde
M}^{R})^{12}|m_{0})(E_{m}^2/4-2)= \nonumber \\
&& [\delta_{m,m_{0}}+
\delta_{m-1,m_{0}}]E^{2}_{m}F^{R}(E_{m})/4  \nonumber \\
&&+(m-1|(\bar{\tilde M}_{R})^{12}|m_{0})+ 
(m+1|(\bar{\tilde M}_{R})^{12}|m_{0})
\end{eqnarray} 
Similar equations can easily be derived for the two remaining blocks. In
the leading order in $m_{0}$ (this approximation is
justified at small voltages $V\rightarrow 0$) at sub-gap energies 
($E_{n}<1$,  $F^{R}=F^{A}=F$)  we obtain
\begin{eqnarray} &&
(n|(\tilde
M^{R})^{11}|m_{0})=\frac{(-1)^{n}(n+1)}{(m_{0}+2)F(E_{m_{0}}^{-})},\\
&& (n|(\tilde
M^{R})^{12}|m_{0})=\frac{(-1)^{n}(n+1)E_{m_{0}}}{(m_{0}+2)E_{n}F(E_{n})},\\
&& (n|(\tilde
M^{R})^{21}|m_{0})=-\frac{(-1)^{n}(n+1)E_{m_{0}}}{(m_{0}+2)E_{n}F(E_{n})}.\\
\nonumber
\end{eqnarray}  
Substituting these matrix elements into eq. (A.7) and performing a simple
summation over $n$ we arrive at the result (\ref{Iar}).

\end{multicols}
\end{document}